
%
\input phyzzx
\catcode`@=11
%
%
\newtoks\UT
\newtoks\monthyear
\Pubnum={UT-\the\UT}
\UT={686}
\monthyear={October, 1994}
\def\p@bblock{\begingroup \tabskip=\hsize minus \hsize
    \baselineskip=1.5\ht\strutbox \topspace-2\baselineskip
    \halign to\hsize{\strut ##\hfil\tabskip=0pt\crcr
    \the\Pubnum\cr
    \the\monthyear\cr }\endgroup}
\def\bftitlestyle#1{\par\begingroup \titleparagraphs
    \iftwelv@\fourteenpoint\else\twelvepoint\fi
    \noindent {\bf #1}\par\endgroup}
\def\title#1{\vskip\frontpageskip \bftitlestyle{#1} \vskip\headskip}
%
%
\def\acknowledge{\par\penalty-100\medskip \spacecheck\sectionminspace
    \line{\hfil ACKNOWLEDGEMENTS\hfil}\nobreak\vskip\headskip}
%
%

%
\def\journal#1&#2(#3){\begingroup \let\journal=\dummyj@urnal
    \unskip, \sl #1\unskip~\bf\ignorespaces #2\rm
    (\afterassignment\j@ur \count255=#3) \endgroup\ignorespaces}
\def\andjournal#1&#2(#3){\begingroup \let\journal=\dummyj@urnal
    \sl #1\unskip~\bf\ignorespaces #2\rm
    (\afterassignment\j@ur \count255=#3) \endgroup\ignorespaces}
\def\andvol&#1(#2){\begingroup \let\journal=\dummyj@urnal
    \bf\ignorespaces #1\rm
    (\afterassignment\j@ur \count255=#2) \endgroup\ignorespaces}

\def\PL{Phys.~Lett.}
\def\PR{Phys.~Rev.}
\def\PRL{Phys.~Rev.~Lett.}
\def\PTP{Prog.~Theor.~Phys.}
\def\ZP{Z.~Phys.}
\catcode`@=12
%

\titlepage

\title{Derivative Expansion
       in Quantum Theory of Gravitation}

\author{Izawa {\twelverm Ken-Iti}
\foot{\rm JSPS Research Fellow.}}
\address{Department of Physics, University of Tokyo \break
                    Tokyo 113, Japan}

\abstract{
The cosmological term
prevents perturbation based on derivative expansion
in Einstein gravity.
We consider quantum theory of gravitation
invariant under
volume-preserving diffeomorphism and Weyl
transformation, which is suitable for derivative expansion.
}

\endpage

\doublespace


\def\c{\varepsilon}
\def\d{\delta}
\def\e{\epsilon}
\def\f{\phi}

\def\h{\theta}
\def\k{\kappa}
\def\l{\lambda}
\def\m{\mu}
\def\n{\nu}

\def\q{\partial}
\def\r{\rho}
\def\s{\sigma}
\def\t{\tau}

\def\v{\varphi}

\def\y{\eta}

\def\L{\Lambda}

\def\j{\tilde g}


\REF\Wud{For a review, J.~Wudka, preprint UCRHEP-T121.}

\REF\Wei{S.~Weinberg \journal Physica &96A (79) 327.}

\REF\Dob{S.~Weinberg, in {\sl General Relativity},
         ed. S.W.~Hawking and W.~Israel
         (Cambridge University Press, 1979); \nextline
         A.~Dobado and A.~L{\'o}pez \journal \PL &B316 (93) 250;
         preprint FT/UCM/13/94; \nextline
         J.F.~Donoghue \journal \PRL &72 (94) 2996;
         preprint UMHEP-408.}

\REF\Buc{W.~Buchm{\"u}ller and N.~Dragon \journal \PL &B207 (88) 292;
         \nextline
         N.~Dragon and M.~Kreuzer \journal \ZP &C41 (88) 485; \nextline
         M.~Kreuzer \journal Class.~Quantum~Grav. &7 (90) 1303.}

\REF\Bij{J.J.~van~der~Bij, H.~van~Dam, and Y.J.~Ng
         \journal Physica &116A (82) 307.}

\REF\Hen{See also, M.~Henneaux and C.~Teitelboim \journal \PL &B222 (89) 195;
         \nextline
         Izawa K.-I. \journal \PTP &91 (94) 393.}

\REF\Kug{S.~Weinberg \journal \PL &9 (64) 357;
         \andjournal \PR &135 (64) B1049; \andvol &138 (65) B988; \nextline
         D.G.~Boulware and S.~Deser \journal Ann.~Phys. &89 (75) 193;
         \nextline
         T.~Kugo and S.~Uehara \journal \PTP &66 (81) 1044.}

\sequentialequations

%
{\caps 1. Introduction}

Universality in quantum field theory
leads to natural expectation that
low-energy physics can be described by effective Lagrangian\rlap.
\refmark{\Wud}
In fact, the standard model
of electroweak and strong interactions may be viewed as
effective field theory
with cutoff stemming from new physics beyond it such as grand unification.
Within perturbation theory, its
renormalizable part provides
dominant contribution to amplitudes computed in the model.

Low-energy description of the other fundamental force
--- graviton on the flat background --- seems slightly different
due to essential non-polynomiality of gravitational field theory.
It is to be compared with chiral Lagrangian in hadron physics,
which is also non-polynomial.
Chiral perturbation theory is based on derivative expansion\rlap.
\refmark{\Wei}

In the case of Einstein gravity without the cosmological term,
derivative expansion in powers of $p/M$
yields effective field theory of graviton\rlap,
\refmark{\Dob}
where $p$ denotes a characteristic momentum
and $M$ is the Planck mass.
However, the cosmological term cannot be suppressed
when a coupled system of gravity and field theory
with spontaneously broken symmetry
is considered.
In order to obtain
the flat background on the true vacuum,
we generically need a non-zero value of the cosmological constant
on the symmetric vacuum.

The cosmological term brings in non-derivative
interaction of gravitons, which prevents perturbation based on
derivative expansion in quantum theory of Einstein gravity.
The order of derivative expansion in four dimensions behaves as
$$
  2L + \sum_n (D_n - 2),
 \eqn\ORDER
$$
where $L$ denotes the number of loops and $D_n$ is the number
of derivatives in each vertex labeled by $n$ in a Feynman diagram.
This indicates troublesome nature of non-derivative interaction
$D_n = 0$ for perturbation based on derivative expansion.

In this paper, we consider quantum theory of gravitation
which is invariant under
volume-preserving diffeomorphism and Weyl transformation.
Weyl invariance excludes the apparent cosmological term from
the effective Lagrangian, which makes perturbation based on
derivative expansion possible
in the theory of gravitation.
Hence this prescription gives a low-energy description
of fundamental forces when it is combined with a model
of other interactions of elementary particles.

%
{\caps 2. Lagrangian}

In order to formulate a theory of massless spin-two particle
in terms of a symmetric tensor field of degree two,
it is necessary to impose gauge symmetries
which eliminate spin-one and spin-zero components in it.
The candidate symmetries are diffeomorphism and Weyl transformation.
The ordinary choice is the full diffeomorphism,
which results in Einstein gravity.
On the other hand, one can adopt volume-preserving diffeomorphism
and Weyl transformation, which is the choice made in this paper.

Let us consider the form of effective Lagrangian
for the metric field $g_{\m \n}$
invariant under
volume-preserving diffeomorphism and Weyl transformation.
For simplicity, we take a scalar $\f$
as an example of matter field.

Weyl symmetry
$$
  \d_{_W} g_{\m \n} = \L g_{\m \n}, \quad \d_{_W} \f = 0
 \eqn\WEYL
$$
implies that the metric field comes in the Lagrangian
solely in the combination
$$
  {\bar g}_{\m \n} = g^{-{1 \over 4}} g_{\m \n},
 \eqn\METRIC
$$
where $g = -\det g_{\m \n}$.

Invariance under volume-preserving diffeomorphism
$$
  \d_{_V} {\bar g}_{\m \n} = -\c^\r \q_\r {\bar g}_{\m \n}
                      - {\bar g}_{\m \r} \q_\n \c^\r
                      - {\bar g}_{\n \r} \q_\m \c^\r, \quad
  \d_{_V} \f = -\c^\m \q_\m \f; \quad
  \q_\m \c^\m = 0
 \eqn\VOL
$$
can be achieved
\refmark{\Buc}
in complete analogy to
the case of full diffeomorphism invariance.
The Lagrangian bears the same form as
that in Einstein gravity except for the field $g_{\m \n}$
in the latter replaced by ${\bar g}_{\m \n}$ in the former\rlap:
\foot{By means of a scalar field $\v$ with a Weyl transformation law
$\d_W \v = -\L \v$, one can construct an action $I[\v g_{\m \n}]$
which classically has both diffeomorphism and Weyl invariances,
where $I[g_{\m \n}]$ denotes a diffeomorphism-invariant action.
A gauge choice $\v = 1$ in $I[\v g_{\m \n}]$
leads to Einstein gravity $I[g_{\m \n}]$, whereas
another choice $\v = g^{-{1 \over 4}}$ results in the present action
with residual symmetries of volume-preserving diffeomorphism and
Weyl transformation.}
$$
  {\cal L} = V(\f) + K(\f){\bar g}^{\m \n}\q_\m \f \q_\n \f
   + U(\f){\bar R} + \cdots;
  \quad U(0) = {1 \over \k^2},
 \eqn\LAGR
$$
where ${\bar g}^{\m \n}$ denotes the inverse
of ${\bar g}_{\m \n}$,
$\bar R$ is the scalar curvature corresponding to ${\bar g}_{\m \n}$,
and the ellipsis stands for higher-derivative terms.
Note that the volume density
(in particular, the apparent cosmological term)
is absent from the expression \LAGR\
since $\det {\bar g}_{\m \n} = -1$.
Hence this theory can serve as a basis for perturbative treatment
based on derivative expansion.

%
{\caps 3. Classical Theory}

In this section, we investigate classical contents of the theory \LAGR.
Let $S[{\bar g}_{\m \n}, \f]$ be the corresponding action,
whose equations of motion
are obtained as
$$
  {\d S \over \d g_{\m \n}}
   = g^{-{1 \over 4}}({\d S \over \d {\bar g}_{\m \n}}
    -{1 \over 4}{\bar g}^{\m \n}{\bar g}_{\r \s}
     {\d S \over \d {\bar g}_{\r \s}}) = 0, \quad
  {\d S \over \d \f} = 0
 \eqn\EOM
$$
by means of the definition \METRIC.

On the other hand, we get an equation
$$
  {\bar D}_\m {\d S \over \d {\bar g}_{\m \n}} = 0
 \eqn\NOET
$$
with the aid of the second equation in \EOM\ and
the Noether identity due to diffeomorphism invariance
of an action $S[g_{\m \n}, \f]$,
where ${\bar D}_\m$ denotes the covariant derivative
corresponding to ${\bar g}_{\m \n}$.
Hence the first equation in \EOM\ yields
$$
  {\bar D}_\m ({\d S \over \d {\bar g}_{\m \n}}
   -{1 \over 4}{\bar g}^{\m \n}{\bar g}_{\r \s}
    {\d S \over \d {\bar g}_{\r \s}})
   = -{1 \over 4}{\bar g}^{\m \n} \q_\m ({\bar g}_{\r \s}
    {\d S \over \d {\bar g}_{\r \s}}) = 0,
 \eqn\INT
$$
which indicates that
$$
  {\bar g}_{\r \s}{\d S \over \d {\bar g}_{\r \s}} = 4\l
 \eqn\COM
$$
is a constant independent of spacetime.

Therefore we conclude that the equations of motion are given by
$$
  {\d S \over \d {\bar g}_{\m \n}} -\l {\bar g}^{\m \n} = 0, \quad
  {\d S \over \d \f} = 0,
 \eqn\FEOM
$$
which are none other than those in Einstein gravity
with a partial gauge-fixing $g = 1$.
This shows that the theory \LAGR\ is classically equivalent
to Einstein gravity with the cosmological constant
as an integration constant determined by an initial condition\rlap.
\foot{We may use the Weyl transformation \WEYL\ to impose
the unimodular condition $g = 1$ in the theory \LAGR.
Then the theory is reduced to the unimodular gravity\rlap,
\refmark{\Bij}
which is known to have the features mentioned above\rlap.
\refmark{\Hen}}

A demand of the flat background $\VEV{g_{\m \n}} = \y_{\m \n}$
(or the unbroken translational invariance)
automatically makes the cosmological constant to be zero on the vacuum.
This enables us to make perturbation based on derivative expansion
in quantum theory of \LAGR.

%
{\caps 4. Quantization}

Quantization of the theory proceeds via
covariant gauge-fixing.
Let us consider the harmonic gauge $\q_\m \j^{\m \n} = 0$,
where we define $\j^{\m \n} = \sqrt{g} g^{\m \n}$.

The BRS transformation $\d$ for the gauge symmetries \WEYL\ and \VOL\
is given by
$$
  \d \j^{\m \n} = \k C \j^{\m \n} - c^\r \q_\r \j^{\m \n}
                 + \j^{\m \r} \q_\r c^\n
                  + \j^{\n \r} \q_\r c^\m, \ \
  \d \f = -c^\m \q_\m \f; \ \
  c^\m = \k \q_\n c^{\n \m}, \ \  c^{\m \n} = -c^{\n \m},
$$
$$
  \d C = -c^\m \q_\m C, \quad
  \d c^{\m \n} = -\k \q_\r c^{\r \m} \q_\s c^{\s \n}
               + i\e^{\m \n \r \s} \q_\r d_\s, \quad
  \d d_\m = \q_\m f,
 \eqn\BRS
$$
where ghosts for ghosts $d_\m$ and $f$
have been introduced and
$\e^{\m \n \r \s}$ denotes the Levi-Civita symbol.
We further introduce additional fields
$$
  \d {\bar C}_\m = iB_\m,
   \quad \d {\bar d}^\m = {\bar c}^\m,
    \quad \d c = id, \quad \d {\bar f} = i{\bar d}
 \eqn\FGH
$$
to adopt gauge conditions
$$
  \k^{-1} \q_\m \j^{\m \n} = 0,
  \quad \e_{\m \n \r \s} \j^{\n \t} \q_\t c^{\r \s} = 0,
  \quad \q_\m {\bar d}^\m = 0,
  \quad \j^{\m \n} \q_\m d_\n = 0,
 \eqn\GFC
$$
which are implemented by a gauge-fixing term
$$
  {\cal L}_B = -i\d ({\bar C}_\m \k^{-1} \q_\n \j^{\n \m}
                     + {\bar d}^\m \e_{\m \n \r \s} \j^{\n \t} \q_\t c^{\r \s}
                      + c \q_\m {\bar d}^\m
                       + {\bar f} \j^{\m \n} \q_\m d_\n).
 \eqn\GFP
$$

With a gauge-fixed Lagrangian ${\cal L}_T = {\cal L} + {\cal L}_B$ in hand,
we can make a perturbative quantization of our theory
on the flat background.

Let us adopt $h^{\m \n}$ as our basic variable, where we define
$$
  \j^{\m \n} = \y^{\m \n} + \k h^{\m \n}.
 \eqn\FLUCT
$$
The quadratic part in the field $h^{\m \n}$ of the gauge-fixed Lagrangian
${\cal L}_T$ is given by
$$
  {\cal L}_0 = {1 \over 4}(\q_\m h_{\n \r} \q^\m h^{\n \r}
                -2\q_\m h^{\m \n} \q^\r h_{\r \n}
                -{3 \over 8}\q_\m h \q^\m h
                +\q_\m h^{\m \n} \q_\n h)
               + B_\m \q_\n h^{\n \m},
 \eqn\QUAD
$$
where $h = \y_{\m \n} h^{\m \n}$.
Hence the corresponding propagator of $h^{\m \n}$
is obtained in momentum space as follows:
$$
  \VEV{h^{\m \n} h^{\r \s}}_0 = {i \over k^2}
   (\h^{\m \r} \h^{\n \s} + \h^{\m \s} \h^{\n \r} - 6\h^{\m \n} \h^{\r \s});
  \quad \h^{\m \n} = \y^{\m \n} - {k^\m k^\n \over k^2},
 \eqn\PROP
$$
which describes desired propergation of massless spin-two particle
on the flat background.

A general theorem
established by Kugo and Uehara
\refmark{\Kug}
in relativistic quantum field theory
states that massless spin-two particle with non-vanishing coupling
in the infrared limit $p_\m \rightarrow 0$
is uniquely the graviton.
Therefore
the present theory provides quantum theory
of gravitation though it is not based on Einstein gravity.
This may be expected from the classical analysis in section 3
that it is essentially equivalent
to Einstein gravity with a partial gauge-fixing.

%
{\caps 5. Conclusion}

We have made a perturbative construction of quantum gravitational theory
on the flat background.
It is based on derivative expansion by means of effective Lagrangian
invariant under volume-preserving diffeomorphism and Weyl transformation.

The construction confirms the view that
quantum theory of gravitation is no less than
the standard model of electroweak and strong interactions
as effective field theory.
In fact, the former may be more accurate than the latter
under the circumstances that
the effective cutoff in the former is expected to be larger
than that in the latter.

Thus one can include the graviton among the participants
in the Standard Model of elementary particles.

%
\acknowledge

The author would like to thank T.~Kugo
for valuable discussions.


\endpage

\refout

\bye